\documentclass[conference]{IEEEtran}
\IEEEoverridecommandlockouts
\usepackage{cite}
\usepackage{amsmath,amssymb,amsfonts}
\usepackage{algorithmic}
\usepackage{graphicx}
\usepackage{textcomp}
\usepackage{xcolor}
\usepackage{import}
\usepackage{url}
\usepackage{enumitem}
\usepackage{multirow}
\usepackage[linesnumbered,ruled,vlined]{algorithm2e}

\def\BibTeX{{\rm B\kern-.05em{\sc i\kern-.025em b}\kern-.08em
    T\kern-.1667em\lower.7ex\hbox{E}\kern-.125emX}}

\SetAlgorithmName{Process}{Process}{List of Processes} 
\begin{document}

\title{Automating a Complete Software Test Process Using LLMs: An Automotive Case Study \\
}

\author{\IEEEauthorblockN{
Shuai Wang\textsuperscript{1}, 
Yinan Yu\textsuperscript{1}, 
Robert Feldt\textsuperscript{1}, 
Dhasarathy Parthasarathy\textsuperscript{2}}
\IEEEauthorblockA{
\textsuperscript{1} Chalmers University of Technology \quad
\textsuperscript{2} Volvo Group}
\IEEEauthorblockA{
Gothenburg, Sweden}
\IEEEauthorblockA{
shuaiwa@chalmers.se, 
yinan@chalmers.se, 
robert.feldt@chalmers.se, 
dhasarathy.parthasarathy@volvo.com}
}

\maketitle

\begin{abstract}
Vehicle API testing verifies whether the interactions between a vehicle's internal systems and external applications meet expectations, ensuring that users can access and control various vehicle functions and data. However, this task is inherently complex, requiring the alignment and coordination of API systems, communication protocols, and even vehicle simulation systems to develop valid test cases. In practical industrial scenarios, inconsistencies, ambiguities, and interdependencies across various documents and system specifications pose significant challenges. This paper presents a system designed for the automated testing of in-vehicle APIs. By clearly defining and segmenting the testing process, we enable Large Language Models (LLMs) to focus on specific tasks, ensuring a stable and controlled testing workflow. Experiments conducted on over 100 APIs demonstrate that our system effectively automates vehicle API testing. The results also confirm that LLMs can efficiently handle mundane tasks requiring human judgment, making them suitable for complete automation in similar industrial contexts. 

\end{abstract}

\begin{IEEEkeywords}
software testing, vehicle API testing, test automation, large language model
\end{IEEEkeywords}

\section{Introduction}
Large Language Models (LLMs) are revolutionizing software engineering. In the past few years, we have witnessed the application of LLMs for assisting or automating numerous software engineering tasks like requirements engineering, software design, coding, and testing \cite{DBLP:conf/fose-ws/FanGHLSYZ23} \cite{DBLP:journals/corr/abs-2308-10620}. Software testing, in particular, is one area where LLMs have been applied with vigor. Facing ever-increasing needs for automation due to the volume and intensity of work involved, testing is rapidly benefiting from the generative capabilities of LLMs. As systematically surveyed in \cite{DBLP:journals/tse/WangHCLWW24}, LLMs have been applied in many testing tasks including system input generation, test case generation, test oracle generation, debugging, and program repair.

While a considerable amount of recent literature has focused on applying LLMs in narrowly scoped tasks \cite{wang2024software} -- such as specific unit tests \cite{chen2023teaching}\hspace{-0.1pt}\cite{yuan2023no}, isolated integration tests~\cite{ajiga2024enhancing}, or individual verification scenarios~\cite{yoon2024intent}\hspace{-0.1pt}\cite{feldt2023towards} -- few have reported on their application to automate a complete test process. Practical testing processes are a diverse mix of steps that are mechanical, creative, and anything in between \cite{fani2023llms}\hspace{-0.1pt}\cite{boukhlif2024llms}. They also involve several (teams of) engineers and tools, whose harmonious cooperation is essential to ensure the quality and cadence of testing. The challenge is only greater when testing automotive embedded systems, where software coexists with mechatronics and other physical systems. Under such heterogeneous conditions, it is not immediately apparent how one can effectively integrate LLMs into a testing process and gain efficiencies. 
In response to these challenges, we present a case study that (1) focuses upon a real-world test process in the automotive industry that is largely performed manually, and (2) automates it using a recipe that seamlessly combines selective use of LLMs with conventional automation.

\begin{figure}[t]
    \centering
    \includegraphics[width=1.0\linewidth]{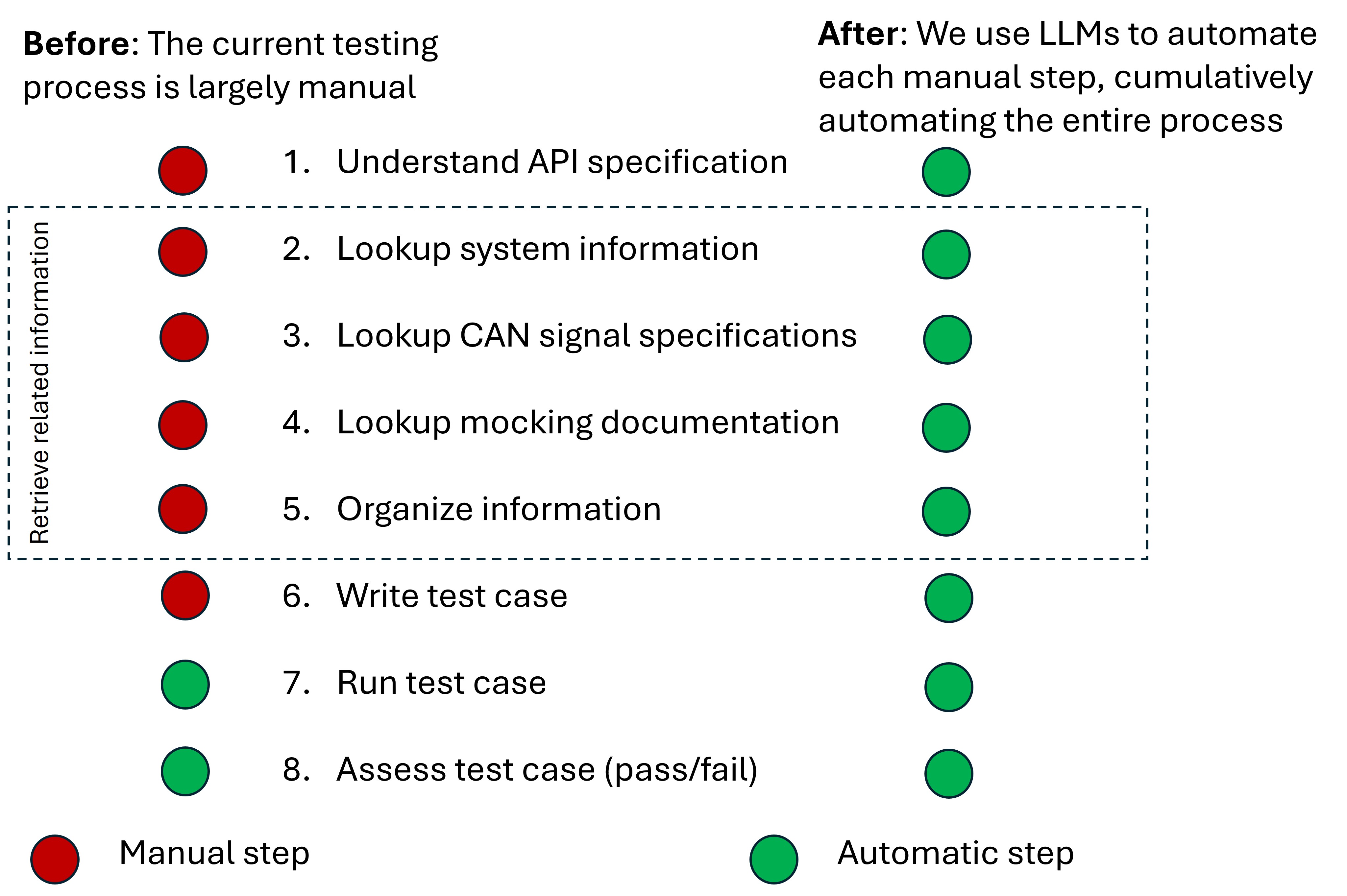}
    \caption{We present the case of automatically testing SPAPI, an in-vehicle web server. Previously, the multistep process of testing SPAPI was largely manual. Using LLMs to automate each manual step, we achieve complete automation.}
    \label{fig:spapi-test-process}
\end{figure}

The focus of this case study -- our system under test -- is SPAPI, a web server that is deployed in trucks made by a leading vehicle manufacturer. SPAPI exposes a set of REST APIs which can be used by clients to read or write selected vehicle states. For example, SPAPI exposes \texttt{/speed} that can be used to read the vehicle speed, and \texttt{/climate} that can be used to change the cabin climate. Essentially, SPAPI serves as a gateway between web clients (like apps on a tablet) on one side, and in-vehicle control and monitoring applications on the other side. More importantly for the purposes of this paper, since SPAPI enables crucial customer-facing applications, considerable effort is spent in ensuring its quality. 

Testing SPAPI requires a dedicated team of 2-3 full-time engineers. As shown in Figure \ref{fig:spapi-test-process} (left), when new APIs are released, the team first reviews the API specifications. They then (2-3) consult multiple documentation sources to understand the associated vehicle states, (4-5) organize this information to determine appropriate mocks and test inputs, and (6-7) write and integrate test cases into a nightly regression suite. Finally, they assess results (8), particularly test failures, to identify valid problems. Notably, as highlighted in Figure 1, most of this process is still performed manually.

These observations prompt the question -- why is such intense manual effort needed to test an arguably simple gateway server? The main reasons are structural. First, as a gateway, SPAPI’s engineering spans multiple teams with overlapping responsibilities. The three core components—the server, vehicle state system, and mocking system—are developed by separate teams, while testing falls to a fourth team that must interpret disparate documentation from each. Second, SPAPI bridges web applications and traditional in-vehicle systems, which differ fundamentally in documentation style. SPAPI APIs are specified in Swagger, making them machine-readable, whereas vehicle states are documented in a mix of natural and formal languages, often requiring human interpretation. Third, SPAPI testers rely heavily on implicit knowledge built over years to manage inconsistencies across systems and teams, leading to highly specialized expertise that intensifies manual effort and complicates team turnover.

In SPAPI testing, the potential for full automation presents two significant benefits: (1) SPAPI testing can be fully automated, increasing the cadence with which APIs can be delivered to customers, and (2) SPAPI testers can be unburdened of their tedious job, allowing their creative talents to be applied elsewhere. Our observations on SPAPI highlight that full automation is not only beneficial but essential under certain test process conditions. 

Specifically, full automation is crucial in scenarios where testers function as a “glue” between tools, systems, and stakeholders in tasks that rely on judgment rather than creativity. Here, automation enhances engineering quality while improving the testers' experience. Additionally, in testing workflows with extensive manual steps, partial automation offers limited gains, reinforcing the need for a comprehensive, all-or-nothing automation approach. Furthermore, when testers navigate legacy processes weighed down by technical debt, partial debt mitigation falls short; complete automation is necessary to address and eliminate debt effectively, benefiting both testers and the organization as a whole.

Recognizing these advantages and the rapid advancements in LLMs for automating manual processes, we explore the central question: can LLMs serve as the key to fully automating a largely manual test process? To address this, we make the following contributions:

\begin{enumerate}
    \item We argue that a test process with clearly decomposed tasks, many of which are executed manually, is a prime candidate for complete automation based on LLMs.
    \item When these criteria are satisfied, we propose a recipe for full automation that involves (a) retaining the test process structure, (b) leveraging LLMs as a general-purpose tool to automate each manual step, and (c) combining LLMs with conventional automation when required.
    \item We present in-vehicle web server testing as a case study, illustrating how a real-world testing process aligns with our criteria and demonstrating its full automation using our proposed recipe.
    \item As the test process structure remains largely intact, we highlight how evaluating the quality of AI-driven automation can be simplified by independently assessing each step where an LLM is applied.
\end{enumerate}

As the following sections will demonstrate, using a real industrial example of in-vehicle embedded software testing, we show that a manual process like SPAPI testing can be fully automated (see Figure \ref{fig:spapi-test-process}) to deliver practical improvements.
\section{Background}
Since SPAPI is a web server that exposes REST APIs, our case study falls within the ambit of API testing \cite{10.1145/3617175}. Aspects of the SPAPI test process are therefore recognizable within the larger universe of API testing, but there are also several case-specific adjustments, which we now highlight. 

\subsection{System architecture}
\begin{figure}[t]
    \centering
    \includegraphics[width=0.95\linewidth]{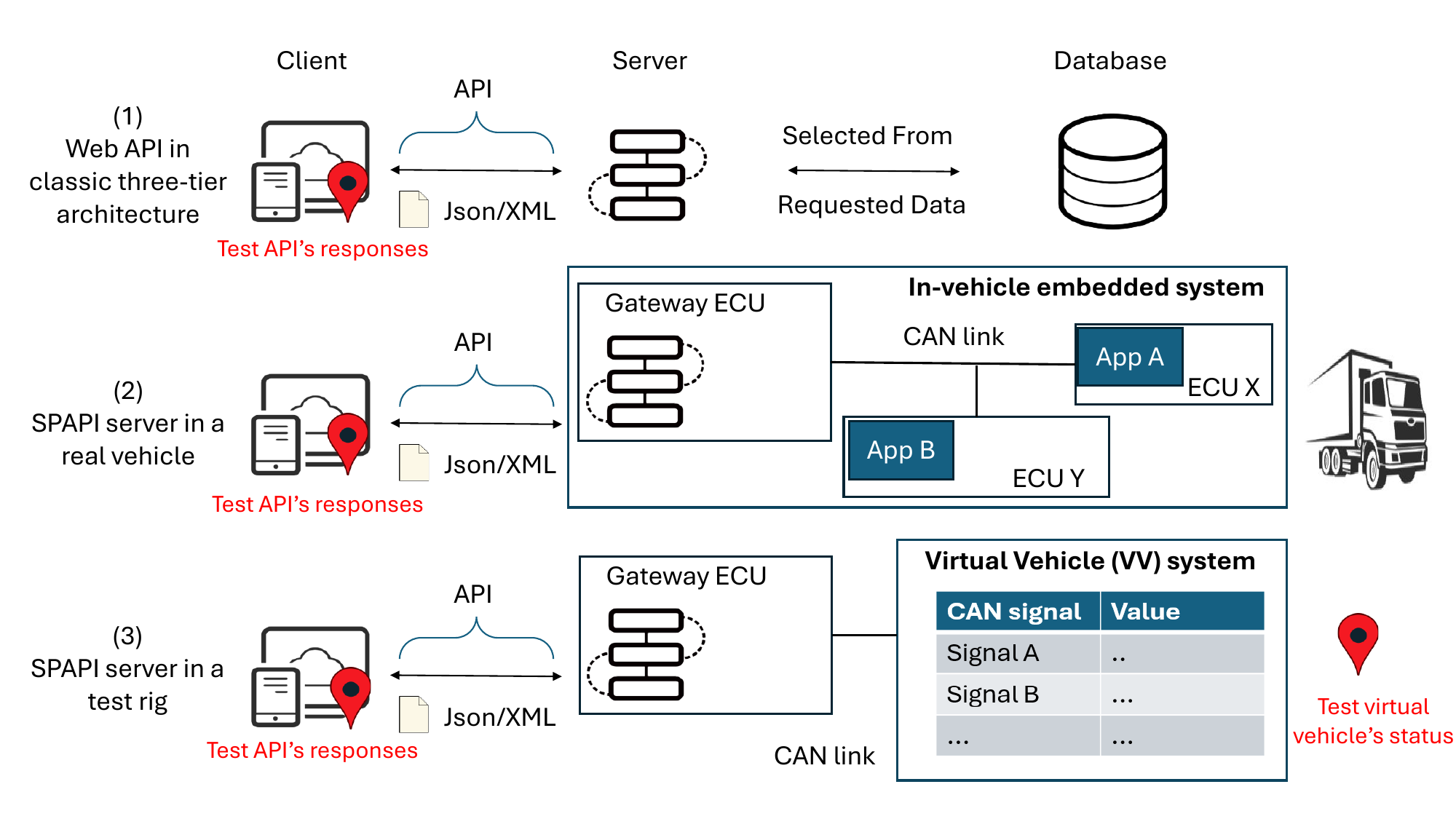}
    \caption{A comparative illustration of the SPAPI architecture -- (1) a web server in the classic three-tier architecture, (2) SPAPI in a real in-vehicle embedded system, and (3) SPAPI in a test rig with vehicle state mocked by a Virtual Vehicle (VV) system. Compared to traditional API testing, vehicle API testing requires not only verifying the API's responses but also checking the vehicle's status.}
    \label{fig:spapi-comparison}
\end{figure}

\begin{figure}[t]
    \centering
    \includegraphics[width=1.0\linewidth]{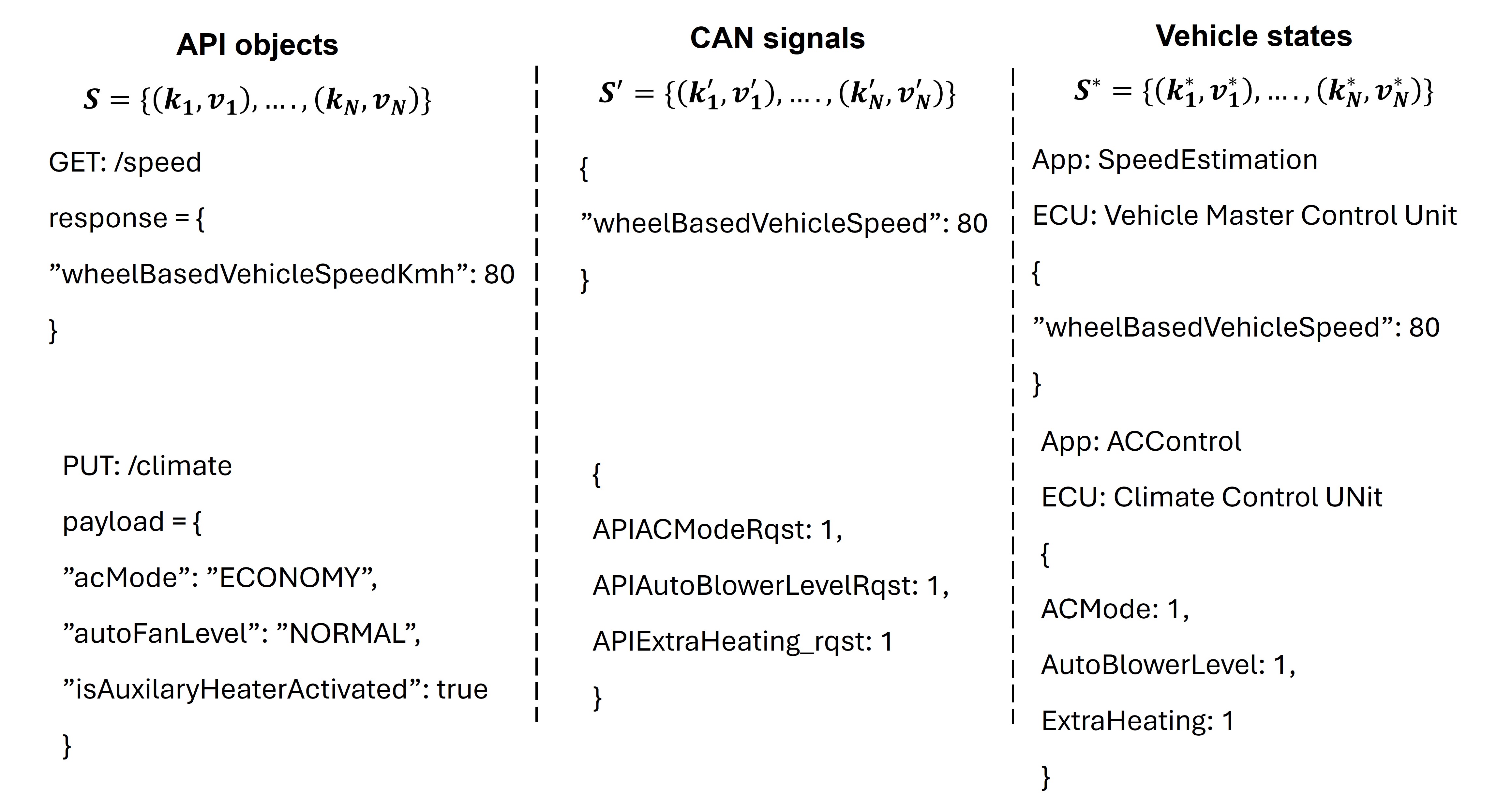}
    \caption{Three tiers of SPAPI operation (1) presentation - SPAPI objects, (2) data access - CAN signals, and (3) data - vehicle states.}
    \vspace{-0.4cm}
    \label{fig:spapi-objects}
\end{figure}

As jointly illustrated in Figures \ref{fig:spapi-comparison} and \ref{fig:spapi-objects}, SPAPI follows the typical 3-tier architecture of decoupling presentation \cite{liu2005modeling}, business logic, and data, each of which we discuss below. 

\noindent \textbf{Presentation} -- Like any web server, SPAPI presents RESTful endpoints with GET and PUT methods and JSON payloads/responses. Each API transacts an object of the form $\mathcal{S} = \{(k_i, v_i)\}_{i=1}^N$, with $N$ attribute-value pairs. Each pair $(k_i, v_i)$ in the object corresponds to some vehicle state $(k^*_i, v^*_i)$ that is managed by a control or monitoring application deployed in an Electronic Control Unit (ECU) in the vehicle. Figure \ref{fig:spapi-objects} shows an example where \texttt{/speed} endpoint provides a GET method that returns the instantaneous speed of the vehicle which, in turn, is calculated by a \texttt{SpeedEstimation} application in a vehicle master control ECU. The same figure also illustrates the \texttt{/climate} endpoint with a PUT method that sets different cabin climate states by communicating with an \texttt{ACControl} application in a climate control ECU. Thus, the essence of SPAPI is presenting APIs for reading or writing an object $\mathcal{S} = \{(k_i, v_i)\}_{i=1}^N$. This corresponds to interacting with vehicle states $\mathcal{S}^* = \{(k^*_i, v^*_i)\}_{i=1}^N$ managed by applications distributed across the in-vehicle embedded system.

\noindent \textbf{Data and data access} -- The typical web server may hold its data in a database, but, clearly, `data' for SPAPI is vehicle state information managed by different in-vehicle control applications. As shown in Figure \ref{fig:spapi-comparison}, these in-vehicle applications are distributed across several ECUs, interconnected using Controller Area Network (CAN) links. While the typical web server may access data by executing database queries, SPAPI accesses data by exchanging CAN signals $\mathcal{S}^\prime = \{(k^\prime_i, v^\prime_i)\}_{i=1}^\mathcal{N}$ with in-vehicle applications. A CAN signal is a pre-defined typed quantity sent through a CAN link between designated sender and receiver applications. In the simplest case, each vehicle state $(k^*_i, v^*_i)$ maps to one CAN signal and value pair $(k^\prime_i, v^\prime_i)$, which SPAPI sends or receives to access the state. We also clarify that this case study focuses upon testing SPAPI in a rig, and not in the real vehicle. In the test rig (see Figure \ref{fig:spapi-comparison}), vehicle state is emulated by a Virtual Vehicle (VV) system, which maintains the superset $\mathcal{N}$ of all vehicle states $\mathcal{S}^* = \{(k^*_i, v^*_i)\}_{i=1}^\mathcal{N}$ in a single table, emulating the state managed by distributed control applications. To maintain consistency of interaction, VV allows state $(k^*_i, v^*_i)$ to be accessed using the same CAN signal $(k^\prime_i, v^\prime_i)$ that SPAPI uses in the real vehicle. In addition to easing testing using virtual means, unlike many other API testing cases, VV offers the advantage of being able to freely mock vehicle state for testing purposes. 
Due to the continuous evolution of CAN signals and the VV platform, it is essential to monitor the vehicle's state to accurately capture relevant state changes.

\looseness=-1

\noindent \textbf{API logic} -- Since SPAPI is a gateway, the logic for each endpoint is relatively lean. When a client invokes an endpoint, SPAPI does the mapping $(k_i, v_i) \rightarrow (k^\prime_i, v^\prime_i)$ of each attribute-value pair in the API object to the corresponding CAN signal-value pair. Then, by sending or receiving the CAN signal and value $(k^\prime_i, v^\prime_i)$, SPAPI reads or writes the corresponding vehicle state $(k^*_i, v^*_i)$. Based upon the result of state manipulation, SPAPI sends an appropriate response to the client. \looseness=-1

\subsection{Current manual API testing}

The current manual workflow for API testing, as shown in Figure~\ref{fig:spapi-test-process}, involves steps such as: understand the API specification, look up related information, write test cases, run and access the test cases. 
Specifically, the tester should first identify the specific object set \( S \) by understanding the documentation. Following this, the tester will retrieve the corresponding CAN signal documentation \( S' \) and the VV system documentation \( S^* \). It is crucial to ensure that each attribute in \( S \) can be mapped to both \( S' \) and \( S^* \). 
This means verifying that every attribute can be converted into a CAN signal and can be simulated in the VV system, and testers can write test cases based on the matched results.
Typically, two key aspects need to be checked during API testing. The first aspect is to verify whether the virtual vehicle's state aligns with expectations after setting certain attributes to specific values via API:
\begin{equation}
  \begin{aligned}
  S^{*} &\leftarrow \text{PUT}(S) \\
  S^{*} &\stackrel{?}{=} S_\text{expected}
\end{aligned}
\end{equation}
The second aspect is to check whether the API returns the expected values under a specific virtual vehicle state:
\begin{equation}
  \begin{aligned}
    S &= \text{GET}() \\
    S &\stackrel{?}{=} S_\text{expected}
\end{aligned}
\end{equation}

In the following content, we will introduce the details of each step. 

\subsubsection{Understand API specification} 
Test engineers need to understand the API documentation to extract the basic objects about the API. The documentation, like Swagger file, always details each API's essential information, such as all available endpoints, expected request formats, and possible response formats for each endpoint. 
Additionally, Swagger defines the data structures used in the API, including objects, properties, and their types.
An example of a Swagger file snippet describing the \textit{Climate} object is shown in Figure~\ref{fig:spapi-tester-workflow}(a).
In this file, testers should parse the object's \textit{acMode} and its corresponding details in the pairs. 
In summary, a thorough understanding the API documentation manually is essential for constructing a comprehensive object set $\mathcal{S} = \{(k_i, v_i)\}_{i=1}^N$ from the original system documentations.

\subsubsection{Retrieve related information}
\begin{figure}[t]
  \centering
  \includegraphics[width=0.9\linewidth]{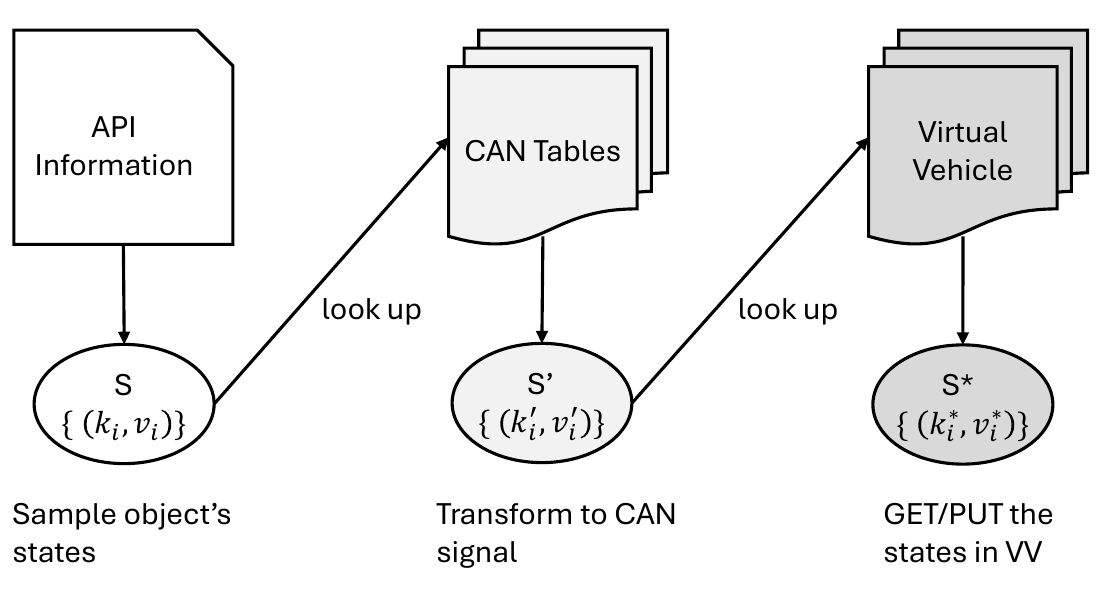}
  \caption{The process of setting and getting vehicle status according to the API information.}
  \vspace{-0.4cm}
  \label{fig:mapping}
\end{figure}

After obtaining the attributes and values corresponding to the object, denoted as $S$, it is necessary to search for related documentation, including the information about CAN signals and the details about the virtual vehicle. The search process is illustrated in Figure~\ref{fig:mapping}.

First, the tester needs to locate the relevant CAN signal documentation from CAN signal table. Then, by matching the corresponding key and value, the original state $S$ is converted into the CAN signal $S'$. Afterward, the relevant virtual vehicle documentation is consulted, and the corresponding key and value are mapped to obtain the specific operation $S^*$ that needs to be performed on the VV.

\textbf{Look up information:}
The three main components of SPAPI testing are the server, vehicle state system, and VV system. Correspondingly, system information, CAN signal specifications and mocking documentation are needed to be retrieved. 

When testing an attribute, the corresponding values should be looked up in both the CAN signal and VV tables. For example, our goal is to set the vehicle's status to \texttt{ECONOMY}.
First, we locate the relevant attribute \texttt{acMode} in the system documentation \( S \). 
Then, we look up \texttt{acMode} in the CAN signal table \( S' \) and find its corresponding value for \texttt{ECONOMY}, which might be 1. 
We then transmit this information to the VV system via the CAN signal. 
Subsequently, in the VV system, we read the corresponding CAN signal and look up the VV table $S^*$ to find the value of the \texttt{acMode} under the \texttt{ECONOMY} state, which might be 2. Finally, we set the value of \texttt{acMode} to 2 in the VV system.
Finally, the \texttt{acMode} in the VV system is set to 2 to achieve the desired vehicle state of \texttt{ECONOMY}.

\textbf{Information organizing:}
In automotive systems, to transmit signals via CAN and utilize VV system correctly, we need to ensure that each attribute and its corresponding value in the system document $\mathcal{S} = \{(k_i, v_i)\}_{i=1}^N$ can be looked up in the CAN signal specifications for getting $\mathcal{S'} = \{(k'_i, v'_i)\}_{i=1}^{N'}$. 
Simultaneously, each attribute and its value in $\mathcal{S'}$ should be looked up in the mocking documentations to get $\mathcal{S^*} = \{(k^*_i, v^*_i)\}_{i=1}^{N^{*}}$. Formally, our goal is to find a mapping such that:
\begin{equation}
  \begin{aligned}
      \forall (k_i, v_i), \;  \exists (k_{j}^{'}, v_j^{'}) \in S^{'} \; &\text{where} \; (k_i, v_i) \rightarrow (k_{j}^{'}, v_j^{'}) \\
      \forall (k_i^{'}, v_i^{'}), \;  \exists (k_{k}^{*}, v_k^{*}) \in S^{*} \; &\text{where} \;  (k_i^{'}, v_i^{'}) \rightarrow (k_{k}^{*}, v_k^{*})
\end{aligned}
\end{equation}
This ensures that every key $k_i$ from $\mathcal{S}$ maps to a corresponding key in $\mathcal{S'}$ and every key $k_i'$ from $\mathcal{S'}$ maps to a corresponding key in \( \mathcal{S^*} \).

However, these three components are developed by different teams, and the corresponding document table may not match exactly, e.g. the names of the attributes in each table may not be consistent since some attributes is recorded using a mixture of natural language and formal language. Table~\ref{table:examples2} summarizes 5 common types of records with different forms. 
Besides, there can be discrepancies in the number of values for an attribute. For example, the \texttt{acMode} attribute may have two states, \texttt{STANDARD} and \texttt{ECONOMY}, in the system document, but there are 3 modes (also \texttt{TURBO}) in the CAN signal specification. In such cases, it is also needed to match the values with equivalent meanings.
Moreover, there are instances of missing attributes, where a corresponding mapping key cannot be found. 
Since such issues are diverse and irregular, testers need to carry out such fuzzy matching cautiously based on their own knowledge and experience.

\begin{table}[h]
  \centering
  \caption{5 types of problems that require fuzzy matching.}
  \begin{tabular}{|c|c|c|}
  \hline
  \multirow{2}{*}{\centering \textbf{Category}} & \multicolumn{2}{|c|}{\textbf{Example}} \\
  \cline{2-3}
   & \textbf{Key 1} & \textbf{Key 2} \\
  \hline
  Spelling errors & \textit{DriverTimeSetting} & \textit{DriverTimeSeting} \\
  \hline
  Abbreviations & \textit{standard} & \textit{STD} \\
  \hline
  Similar writing formats & \textit{standard\_mode} & \textit{STANDARDMODE} \\
  \hline
  Logical equivalents & \textit{OFF} & \textit{NOT\_ON} \\
  \hline
  Semantic equivalents & \textit{AutoStart} & \textit{AutoLaunch} \\
  \hline
  \end{tabular}
  \label{table:examples2}
\end{table}

\subsubsection{Write Test Cases}
Based on the organized information, testers can write reasonable and comprehensive test cases. 
Specifically, the two main methods of a vehicle API, PUT and GET, need to be tested separately. The PUT method is used to set the car's state, while the GET method is used to retrieve the car's current state.
To verify the effectiveness of the PUT method, we set the car's state to \( S \) using the PUT method and then check whether all the virtual vehicle's states $S^{*}$ in the VV system are as expected.

To verify whether the GET method is valid, we directly call the GET method to retrieve the car's current states, and check if the retrieved states \( S \) match the expected states.

The process of writing test cases requires testers to have a comprehensive understanding of the organized information and a background in computer science, such as ensuring the correctness of data types in test cases. In addition, testers need to consider all test situations consider as many test situations as possible to ensure high coverage of test cases.
Finally, testers write the test code to execute the test cases.

\subsubsection{Running code and Evaluating results}
Once the environment and code are prepared, the code can be executed to automatically test the API. Existing test frameworks and tools can be used to organize the results, allowing to directly obtain the final outcomes. 

\subsection{Obstacles to automation}
Based on the current API testing process, the obstacles to achieving automated API testing can be summarized as follows:

\begin{itemize}
  \item \textbf{Fuzzy Matching:} Since the system information, CAN bus specifications and VV documentations are recorded by different teams, the names of the attributes (i.e., the key in the table) are sometimes inconsistent, as shown in Table~\ref{table:examples2}. 
  In addition, the value also needs to be mapped based on the semantics of the key.
  For instance, the attribute \texttt{isAlarmActive} may be \texttt{TRUE}/\texttt{FALSE} in system files but \texttt{Active}/\texttt{Inactive} in CAN specifications. 
  Such inconsistency makes it difficult to achieve exact matching, necessitating the implementation of a fuzzy matching mechanism.
  \item \textbf{Informal Pseudocoded Mappings:} 
  In the CAN signal table, data is often represented in the form of informal pseudocode, leading to situations where a single key-value pair maps to multiple counterparts. For example, activating the car's alarm clock requires setting the attribute and value as \{AlarmActive:True\}. However, the corresponding data in the CAN signal table could be represented as \{AlarmClockStat:Active OR AlarmClockStat:Ringing OR AlarmClockStat:Snoozed\}.
  In this scenario, it is necessary not only to parse the CAN signal table but also to match the original key-value pair with each entry in the CAN signal table. This requires recognizing and parsing these pseudocode forms and being able to handle one-to-many mappings.
  \item \textbf{Inconsistent Units:} Automotive values are often associated with units, such as speed, which can be measured in \texttt{km/h} or \texttt{m/s}. When units are inconsistent, direct mapping of values between tables is not possible. Values must be converted to the corresponding units before mapping. Thus, the variety of units and the different conversions required between them make detecting unit inconsistencies and performing conversions a major challenge.
  \item \textbf{Inter-Parameter Dependencies:} Parameters often have complex interdependencies, requiring coordinated settings. For example, the attribute \texttt{alarmTime} might be represented as a date-time string in system files, but in CAN files, it might need to be mapped separately to \texttt{hours} and \texttt{minutes}. Capturing and managing these parameter relationships is not an easy task.
\end{itemize}

\section{Fully automated SPAPI testing with LLMs}
\begin{figure*}[ht!]
    \centering
    \includegraphics[width=1\textwidth]{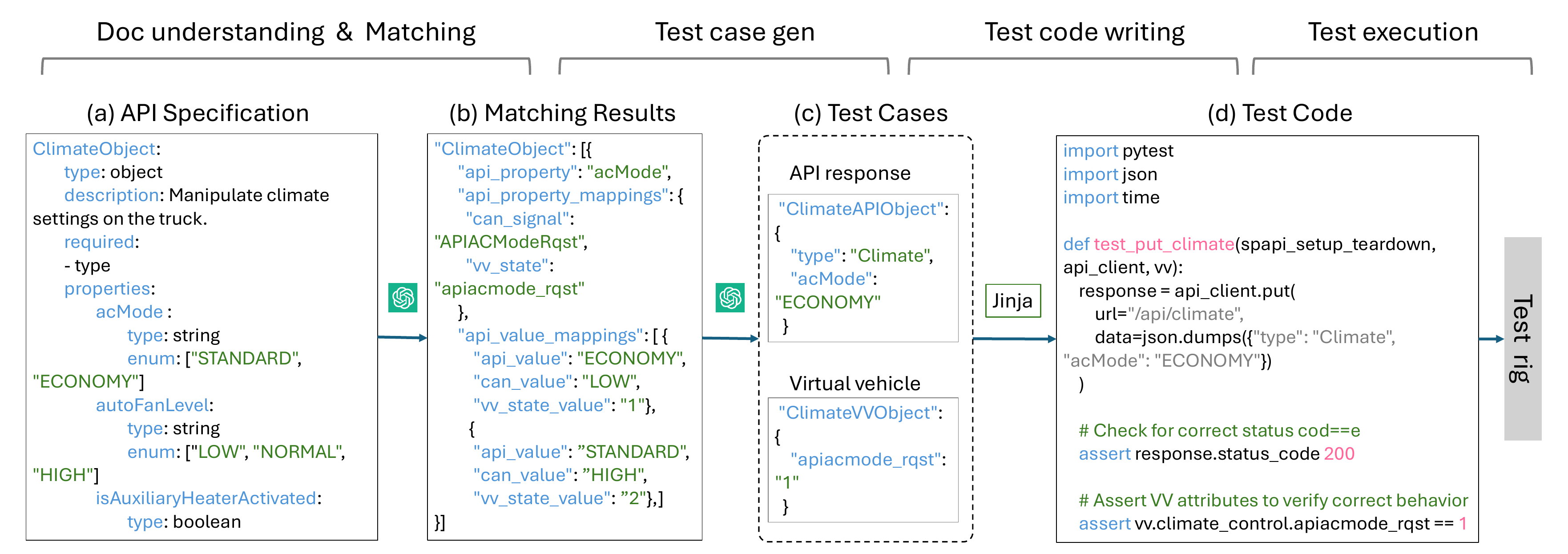}
    \caption{Architecture and workflow of SPAPI-Tester: The pipeline largely preserves the manual process and selectively uses LLMs to automate discrete steps.}
    \vspace{-5mm}
    \label{fig:spapi-tester-workflow}
\end{figure*}

This section presents the details of our automated testing tool, SPAPI-Tester, which can integrate with LLMs to fully automate the entire API testing process.
The overall process, as shown in Figure \ref{fig:spapi-tester-workflow}, can be divided into four main steps. These steps are detailed in Process \ref{alg:spapi_tester_workflow}.

\begin{algorithm}[htbp]
    \caption{Overall Workflow of SPAPI-Tester} \label{alg:spapi_tester_workflow}
    TestTracker = InitializeTestTracker() \\
    \textbf{For} APISpec \textbf{in} List(APISecifications) \\
    \hspace*{2em} S = ExtractTestObjects(APISpec) \\
    \hspace*{2em} S' = APIToCANMapping(S, CANTable) \\
    \hspace*{2em} S* = CANToVVMapping(S', VVTable) \\
    \hspace*{2em} TestCases = GenerateTestCase(S, S*) \\
    \hspace*{2em} TestCode = WritingTestCode(TestCases) \\    
    \hspace*{2em} TestTracker.analyzeTestRun(TestCode) \\
TestReport = PushToTestRepo(TestTracker)
\end{algorithm}

After initializing SPAPI (\textit{line 1}), the entire testing process is divided into four parts:
\textbf{(1)} Documentation understanding (\textit{line 3}): This part involves identifying test objects based on the API specifications. 
\textbf{(2)} Information matching (\textit{lines 4, 5}): This part entails look up relevant CAN table and virtual vehicle documents to matching all these objects. 
\textbf{(3)} Test case generation (\textit{line 6}): Using the matched data, this step focuses on generating test cases for the API's return results and verifying the virtual vehicle's status.
\textbf{(4)} Executing test cases and generating test reports (\textit{line 7, 8, 9}).

\subsection{Documentation understanding}
The purpose of documentation understanding is to extract the test objects from the API documentation.
Standard API documentation, commonly in YAML or Json format~\cite{openapi2023}, as shown in Figure \ref{fig:spapi-tester-workflow}(a), is structured to list attributes and values associated with various objects. This structured format lends itself well to template-based parsing.
We parse these documents and use predefined templates to extract the relevant attributes and values. Based on existing templates~\cite{openapitemplate2024}, we define a few simple and common rules to ensure the method's general applicability.
These templates focus on fundamental elements, such as endpoint names, attribute names, and data types. Additionally, if sample API calls are provided in the documentation, we extract these directly to test the basic accessibility and functionality of the API.

However, using templates alone is insufficient for determining reasonable attribute values. We have identified the following issues with relying only on templates:

(1) Cannot utilize attribute description: API documentation often includes natural language descriptions of attributes that templates cannot interpret or utilize. These descriptions typically contain constraints on the attributes, which are crucial to prevent generating incorrect values.

(2) Lack of robustness: API documentation can sometimes be informal or inconsistent. For instance, attributes of enumeration types are usually presented as ["STANDARD", "ECONOMY"], but some documents might incorrectly use "STANDARD or ECONOMY". Only using templates makes it difficult to address these random and informal issues effectively.

To overcome these two issues, we introduce LLMs to enhance the process.
LLMs are utilized to analyze the entire API documentation, leveraging natural language descriptions to understand attribute constraints more effectively. Since LLMs are capable of semantic understanding, they also mitigate the impact of informal formatting or inconsistencies. This allows the system not only to parse API properties but also to map them to CAN signals, which is covered in detail in the subsequent section. 
LLMs further generate constraints based on attribute descriptions, producing reasonable values within these constraints. The contextual insights provided by LLMs help create a broader set of valid test values, thereby improving the coverage and reliability of our test cases.

\begin{figure}[ht]
    \centering
    \includegraphics[width=1.0\linewidth]{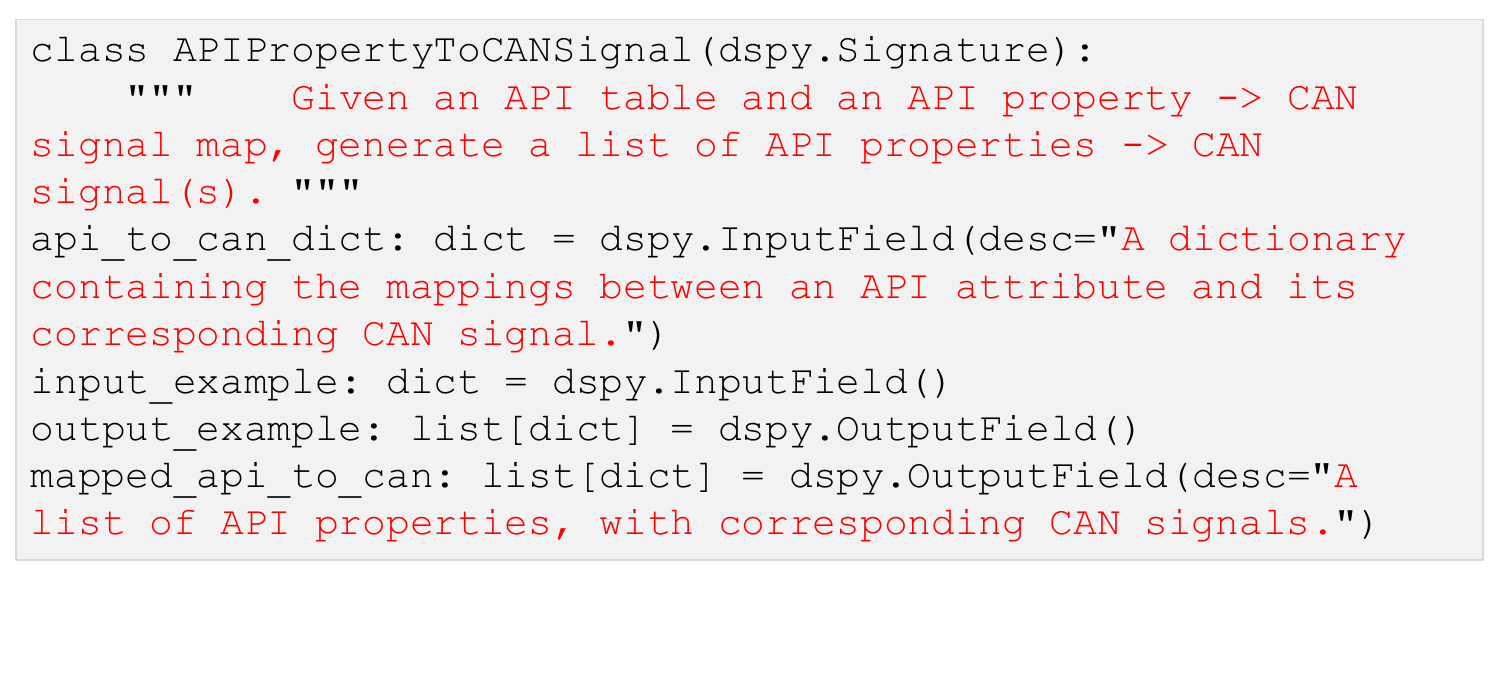}
    \vspace{-1cm}
    \caption{A DSPy Signature for automating API to CAN lookup (simplified).}
    \label{fig:DSPy-Signature}
\end{figure}

In practice, to ensure the stability of LLM outputs and reduce the effect of the specific prompt formulations, we employ DSPy \cite{khattab2023dspy} to automate prompt optimization. 
DSPy enables us to write declarative LLM invocations as Python code. Figure \ref{fig:DSPy-Signature} illustrates a simplified example of one of our prompts, along with the DSPy Signature.
This \texttt{APIPropertyToCANSignal} signature outlines the process of converting structured API properties to CAN signals, which automates the time-consuming task of constructing an API property $(k_i, v_i)$ and mapping it to a corresponding CAN signal $(k'_i, v'_i)$.

\begin{figure}[t]
    \centering
    \includegraphics[width=1.0\linewidth]{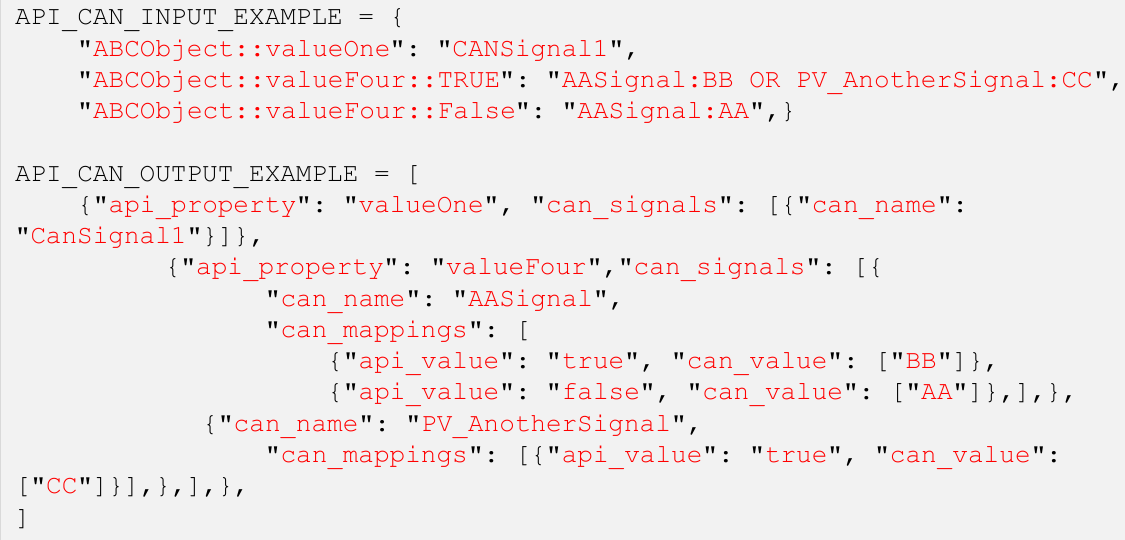}
    \caption{Templatized examples for guiding API to CAN look up.}
    \label{fig:template-mapping}
\end{figure}

To further improve the accuracy and ease of extracting structured data from the LLM, we format the LLM inputs and outputs as dictionaries. We define dictionary-based prompt templates to make tasks more comprehensible for the LLM \cite{openai2023structuredoutputs}, as demonstrated in Figure \ref{fig:template-mapping}. By specifying the expected output fields, the signature directs the LLM to navigate inconsistencies in documentation and accurately associate API properties with CAN signal values. Furthermore, by typing fields in the signature, we enable the use of a \texttt{TypedPredictor} in DSPy, which validates the LLM response. If the response does not conform to the specified types, DSPy re-prompts the LLM, repeating this up to a maximum threshold until compliance is achieved. This structured approach capitalizes on the improved format adherence of LLMs, enhancing consistency and reliability.

\subsection{Information Matching}

As illustrated in Figure \ref{fig:mapping}, the mapping of information in our system encompasses two stages: mapping API properties to CAN signals and mapping CAN signals to Virtual Vehicle (VV) signals. These mappings are crucial for enabling signal transmission within the vehicle as well as setting or verifying the vehicle’s state. Since the processes and methods for these two mappings are similar, we will detail the approach for mapping API properties to CAN signals as an example.

First, we retrieve a set of candidate CAN signal key-value pairs \(\{(k'_i, v'_i)\}\) from a CAN signal library through solely matching the name of endpoint. Subsequently, we use the extracted API attributes \(\mathcal{S} = \{(k_i, v_i)\}_{i=1}^N\) and the candidate CAN signals \(\{(k'_i, v'_i)\}\) as input to an LLM, enabling many-to-many matching between API properties and CAN signals. In many cases, attributes may have multiple enumerated values. For instance, as shown in Figure \ref{fig:template-mapping}, an API property `valueFour' might take the values `True' or `False', while the corresponding CAN signal might represent these states as `AA' and `BB'. This type of mapping is common, and to increase the stability of SPAPI-Tester, we utilize a separate DSPy module specifically for matching enumerated values. The input consists of enumerated values from both the API property and the CAN signal, and the output is a mapping of these values.

As discussed in Section II.C, there are several challenges in the mapping process. First, for \textit{fuzzy matching}, the LLM’s strong semantic understanding is well-suited to handle these cases. Second, for \textit{pseudocode mappings}, we enhance template robustness by embedding examples directly into the prompt, as shown in Figure 8. For example, we map ``AAsignal:BB OR PV\_AnotherSignal:CC" to ``can\_value" : ``BB", thereby minimizing document noise while extracting relevant information. Third, for \textit{unit inconsistencies}, we apply a dedicated DSPy module that uses a Chain-of-Thought (CoT) approach to extract and normalize units within values. This module converts units (e.g., `kW' to `Kilowatts'), ensuring unit alignment in the test case generation phase.

The final output is structured as a list, as defined in Figure \ref{fig:DSPy-Signature}, with each element containing a fully matched pair. The same approach is then applied to map information between CAN signals and VV signals, ultimately yielding complete matching results \(\mathcal{S'}\) and \(\mathcal{S*}\).

\subsection{Test case generation}
\begin{figure}[bh]
    \centering
    \vspace{-0.3cm}
    \includegraphics[width=1.0\linewidth]{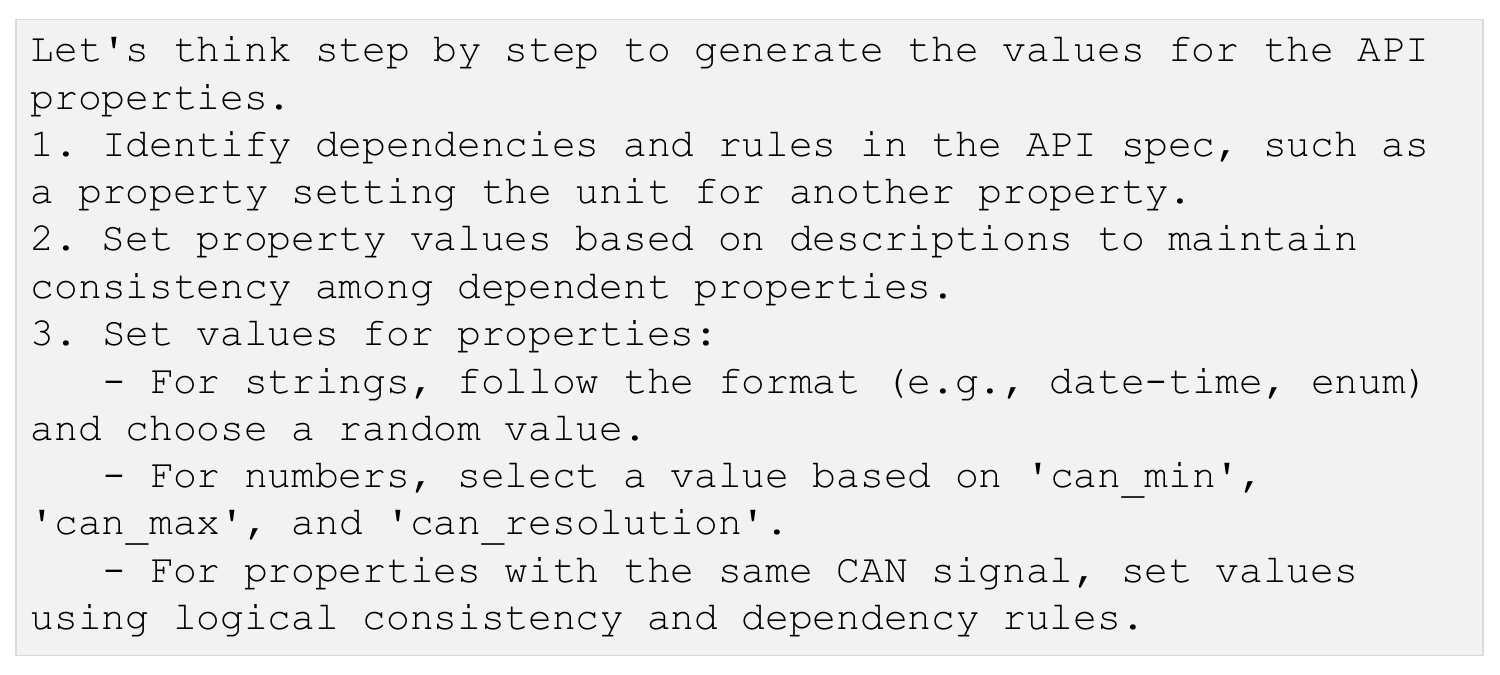}
    \vspace{-0.5cm}
    \caption{Chain-of-Thought prompt for test case generation (simplified).}
    \label{fig:test-case-gen-prompt}
    
\end{figure}

After matching all the necessary information, we integrate these details into a structured document, as illustrated in Figure \ref{fig:spapi-tester-workflow}(b), which then serves as the basis for generating test cases.

Given the need to address multiple constraints during test case generation—such as unit consistency—we employ a stepwise CoT approach to progressively incorporate these constraints. Specifically, for \textit{inconsistent units}, we prompt the LLM to identify relationships between units and perform any necessary conversions. For \textit{inter-parameter dependencies}, the LLM captures relationships among parameters, ensuring compatibility and avoiding value conflicts. Additionally, the LLM identifies property types and manages specialized formats, such as date-time strings. Finally, we guide the LLM in handling cases common in industrial contexts, such as shared CAN signals among multiple properties or specific constraints on value ranges.

To ensure these constraints are applied consistently, we leverage DSPy’s \texttt{TypedChainOfThought} method, which consolidates all conditions within a single prompt. Figure \ref{fig:test-case-gen-prompt} provides a simplified example of this prompt. For ease of use, we specify that the module outputs test cases in dictionary format, as depicted in Figure \ref{fig:spapi-tester-workflow}(c).

After generating the test cases, we use them to create test code.
The test code generally consists of two sections: a setup section, which includes essential elements such as package imports and requests to enable program execution, and a validation section containing assertions.
Since the setup code remains consistent across tests, we design distinct \texttt{Jinja}\footnote{\url{https://palletsprojects.com/projects/jinja/}} templates for PUT and GET test cases.
Using a simple code renderer, we inject the generated API and VV test objects into the \texttt{Jinja} template to render the \texttt{Pytest} test case. Figure \ref{fig:spapi-tester-workflow}(d) shows an example test case rendered by the test-writing module.

\subsection{Executing test cases and generating test reports}
To ensure the automation of the entire process, the system automatically executes the test code on the test rig \cite{asyraf2019fundamentals}, and then generates a comprehensive test report. This report documents the details of the automated testing process, including the test objects, the matching results, the generated test cases, and the execution logs. Such documentation ensures that our system maintains a high level of transparency, rather than functioning as a black box.

\vspace{0.2cm}
\section{Experiments}
Our evaluation investigates the following questions.

\begin{enumerate}[label=RQ\arabic*:]
    \item What are the pass rate, coverage, and failure-detection capability of the test cases generated by the SPAPI-Tester?
    \item To what extent can LLMs overcome the obstacles outlined in Section II.C to achieve end-to-end automated testing?
    \item How efficient is this automated API testing?
    \item How effective is SPAPI-Tester in testing real-world industrial APIs?
\end{enumerate}
Specifically, RQ1-RQ3 focus on ablation studies of SPAPI-Tester, using controlled experiments to evaluate its capabilities and performance. RQ4 examines the application of SPAPI-Tester in the real-world, industrial setting with newly developed (and thus guaranteed to be unseen) APIs to demonstrate the effectiveness of our end-to-end automated testing system.

\subsection{Experimental Setup} 
In this section, we describe our experimental setup. 
\subsubsection{Subjects}
Our research focuses on automating vehicle API testing within an industrial setting, addressing unique challenges such as inconsistencies across documentation and system specifications. As no existing methods directly address these issues in vehicle API automation, we could not compare our approach with general API testing techniques, as they lack the capability to handle the specific requirements of our industrial setting.

We evaluated the quality of generated test cases for 41 truck APIs using metrics such as pass rate and coverage. To assess SPAPI-Tester's error detection capabilities, we annotated an additional 109 APIs developed by a leading vehicle manufacturer. These APIs were supported by system documentation from in-house truck experts, CAN signal protocols from the CAN-bus team, and virtual vehicle documentation from the Virtual Vehicle team.

We tested four LLMs: two classic models—GPT-3.5-turbo (OpenAI, 2023-07-01-preview) and LLaMA3-70B (2024-04-18)—and two recent advancements, GPT-4o (2024-05-13) and LLaMA3.1-70B (2024-07-23). 
To ensure flexibility and reduce maintenance, we opted not to fine-tune these models with company-specific data, allowing seamless adaptation to new models or data without retraining.

\subsubsection{Metrics}
We evaluate our SPAPI-Tester both at the API level and at the test case level.
At API level, we use the \textit{pass rate} of the APIs as our metric. If all generated test cases for a given API pass the tests, we consider that API to have passed. Conversely, if any test case fails, the API is considered to have failed. Therefore, the \textit{pass rate} is defined as the proportion of APIs that pass the tests.

At test case level, we primarily assess the quality and coverage of the generated test cases. For these evaluations, we employ precision and recall as our key metrics. Precision measures the quality of the test cases generated, while recall measures their coverage of API properties.

\begingroup
\setlength{\tabcolsep}{3pt}
\begin{table}[t]
    \centering
    \caption{Pass rate on different types of APIs.}
    \resizebox{0.5\textwidth}{!}{
        \begin{tabular}{|c|c|cccc|}
        \hline
        \multirow{2}{*}{{API Type}} & \multirow{2}{*}{Num.} & \multicolumn{4}{c|}{LLMs} \\
        \cline{3-6}
        & & {GPT-3.5} & {LLaMA3} & {LLaMA3.1} & {GPT-4o} \\
        \hline
        Energy & 8 & 0.88 & 1.0 & 0.88 & 1.0 \\
        Driver Settings & 6 & 0.83 & 0.83 & 1.0 & 0.83 \\
        Visibility Control & 11 & 0.91 & 1.0 & 0.91 & 1.0 \\
        Software Control & 3 & 1.0 & 1.0 & 1.0 & 1.0 \\
        Vehicle Condition & 9 & 1.0 & 1.0 & 1.0 & 1.0 \\
        Other & 4 & 1.0 & 1.0 & 1.0 & 1.0 \\
        \hline
        Total/Average & 41 & 0.93 & 0.98 & 0.95 & 0.98 \\
        \hline
        \end{tabular}
    }
    \label{table:main_results}
    \vspace{-0.5cm}
\end{table}
\endgroup

\subsection{Pass Rate, Coverage, and Failure Detection (RQ1)}
\textit{Pass rate:} Since APIs with similar functions typically call the same electronic control unit (ECU) in embedded systems and, thus, share documentation within the same domain, we grouped 41 truck APIs into 6 categories based on their functions to present the results more clearly. Table~\ref{table:main_results} details the pass rates for each category.

These 41 APIs are online and pre-verified, ensuring that any failures observed during testing were due to issues within the generated test cases or code. Results show that for the majority of categories, all the APIs can pass the tests successfully, with all four LLMs achieving high pass rates. Notably, SPAPI-Tester achieved a 98\% pass rate when using LLaMA3 and GPT-4o, demonstrating the method's accuracy in generating valid test samples. However, GPT-3.5 exhibited slightly lower performance in handling structured input-output, failing in two cases due to improper CAN connection settings. Additionally, a common error across all LLMs stemmed from missing unit descriptions in API specifications. For example, when documentation omitted units for battery power, LLMs incorrectly defaulted to watts (W) instead of kilowatts (kW), leading to test case failures. Broad patterns of errors like this could likely be addressed by further refining the prompts.

\begin{table*}[h!]
    \centering
    \caption{Test case coverage of different types of APIs. 'P' is Precision, 'R' is Recall, and 'F1' is the F1 score.}
        \begin{tabular}{|c|c|c|c|c|c|c|c|c|c|c|c|c|}
            \hline
            \multirow{2}{*}{\centering API Type} & \multicolumn{3}{|c|}{GPT-3.5} & \multicolumn{3}{|c|}{LLaMA3} & \multicolumn{3}{|c|}{LLaMA3.1} & \multicolumn{3}{|c|}{GPT-4o} \\

            \cline{2-13}
             & \textbf{P} & \textbf{R} & \textbf{F1} & \textbf{P} & \textbf{R} & \textbf{F1} & \textbf{P} & \textbf{R} & \textbf{F1} & \textbf{P} & \textbf{R} & \textbf{F1}\\
            \hline
            Energy & 0.96 & 0.69 & 0.78 & 0.98 & 0.76 & 0.85 & 0.96 & 0.74 & 0.84 & 0.96 & 0.79 & 0.87\\
            Visibility Control & 0.97 & 0.70 & 0.78 & 0.96 & 0.70 & 0.79 & 0.97 & 0.74 & 0.84 & 0.96 & 0.80 & 0.87\\
            Vehicle Condition & 1.0 & 0.95 & 0.97 & 1.0 & 0.9 & 0.95 & 1.0 & 0.95 & 0.97 & 1.0 & 0.95 & 0.97\\
            Other & 1.0 & 0.63 & 0.77 & 1.0 & 0.85 & 0.92 & 1.0 & 0.83 & 0.91 & 1.0 & 0.80 & 0.89\\
            \hline
            \textbf{Average} & \textbf{0.97} & \textbf{0.73} & \textbf{0.80} & \textbf{0.98} & \textbf{0.79} & \textbf{0.88} & \textbf{0.98} & \textbf{0.81} & \textbf{0.89} & \textbf{0.97} & \textbf{0.85} & \textbf{0.90}\\
            \hline
        \end{tabular}
    \label{table:performance_metrics}
    \vspace{-0.3cm}
\end{table*}

\textit{Coverage:}
In addition to pass rate analysis, we evaluated the coverage of generated test cases to assess whether they adequately test each API. A vehicle expert group was invited to create ground truth test cases for 12 representative APIs, each including 5 to 30 test cases across 4 categories. The results are presented in Table~\ref{table:performance_metrics}. All LLMs demonstrated high precision, with precision rates exceeding 0.97 across the board and reaching 1.0 for half of the APIs, showcasing the high quality of test cases generated by our model. For cases where precision was below perfect, errors originated from limitations in the fuzzy matching step.

However, recall rates did not reach optimal levels primarily due to missing information in the API documentation, such as absent units or variable types for some attributes. To maintain high precision, SPAPI-Tester skips samples that lack sufficient context for accurate matching, resulting in a recall loss of approximately 15 percentage points. All untested attributes are logged in the testing report, allowing developers to trace and address these underlying issues.

\textit{Failure detection:}
To further assess the effectiveness of the generated test cases in detecting failures, vehicle experts labeled 109 additional truck APIs, being developed, identifying 38 as buggy. SPAPI-Tester created test cases that successfully detected all buggy APIs with only four false positives, achieving a 96\% accuracy rate. 

All models performed comparably, highlighting that our stepwise, structured pipeline design reduces dependence on specific LLM choices.
We seamlessly migrated SPAPI-Tester to different LLMs without requiring additional adaptation. This largely model-agnostic pipeline design allows us to focus on refining the testing process rather than selecting specific LLMs, given the abundance of options.

\subsection{LLMs' ability of overcoming obstacles  (RQ2)}
\textbf{Fuzzy matching} presents a significant challenge in automated API testing. We categorized common fuzzy matching examples into five classes, selecting 20 test samples per class, supplementing with manually written samples if needed. 
The results, shown in Table~\ref{table:fuzzy_matching} (upper part), indicate that all models achieved high precision rates, highlighting the LLMs' capability to accurately recognize and match fuzzy inputs, a key requirement for full automation. For \textit{semantic equivalents, logical equivalents, and similar writing formats}, all the models attained an accuracy of 1.0 or nearly so, demonstrating their strong pattern matching abilities in semantics and logic. However, for \textit{spelling errors}, accuracy slightly dropped as some errors altered word semantics, like mistaking \texttt{date} for \texttt{data}. In the \textit{abbreviations} category, some abbreviations were too short to discern, complicating the matching process. 

For the \textbf{inconsistent units} issue, we selected 200 samples for experiment. The results in Table~\ref{table:fuzzy_matching} (lower part) indicate that while SPAPI-Tester achieves a high precision rate, the recall remains suboptimal. The reason is that some documentation explicitly annotates units for each attribute, while others omit these details. In these cases, it becomes necessary to infer the units based on descriptions or other contextual information, which can affect the performance.

\begin{figure}[t]
    \centering
    \includegraphics[width=0.8\linewidth]{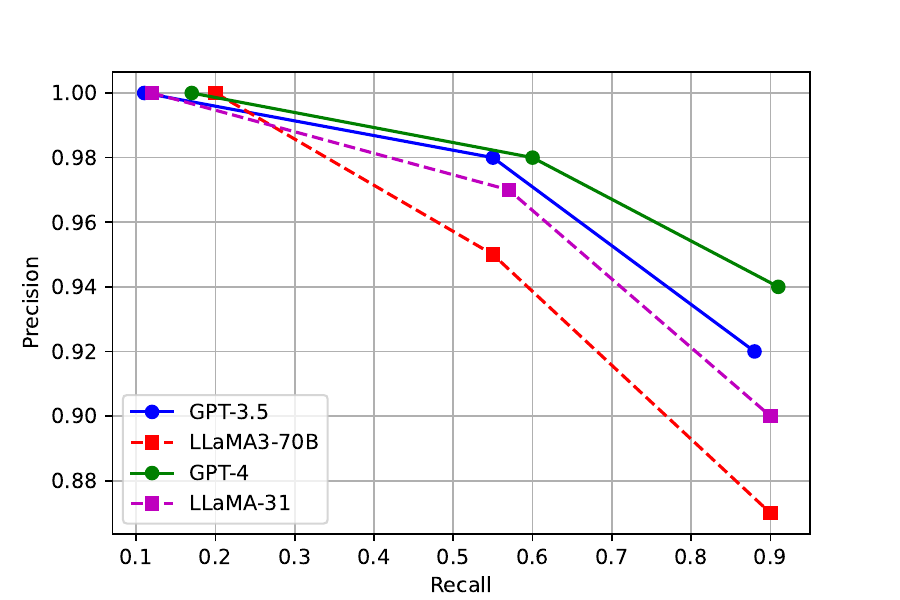}
    \caption{Matching performance on informal pseudocoded mappings.}
    \label{fig:ambiguous}
    \vspace{-0.4cm}
\end{figure}

\begin{table*}[h]
    \centering
    \caption{Performance on different types of Fuzzy Matching (upper part) and Inconsistent Units (lower part).}
        \begin{tabular}{|c|c|c|c|c|c|c|c|c|c|c|c|c|}
            \hline
            \multirow{2}{*}{\centering API Type} & \multicolumn{3}{|c|}{GPT-3.5} & \multicolumn{3}{|c|}{LLaMA3} & \multicolumn{3}{|c|}{LLaMA3.1} & \multicolumn{3}{|c|}{GPT-4o} \\

            \cline{2-13}
             & \textbf{P} & \textbf{R} & \textbf{F1} & \textbf{P} & \textbf{R} & \textbf{F1} & \textbf{P} & \textbf{R} & \textbf{F1} & \textbf{P} & \textbf{R} & \textbf{F1}\\
    \hline
    Spelling errors & 0.89 & 0.76 & 0.82 & 0.92 & 0.73 & 0.81 & 0.91 & 0.78 & 0.84 & 0.91 & 0.83 & 0.87\\
    Abbreviations & 0.93 & 0.68 & 0.79 & 0.88 & 0.74 & 0.80 & 0.92 & 0.75 & 0.83 & 0.98 & 0.74 & 0.84\\
    Similar writing formats & 0.95 & 0.95 & 0.95 & 1.0 & 0.95 & 0.97 & 1.0 & 0.95 & 0.97 & 0.95 & 0.95 & 0.95\\
    Logical equivalents & 1.0 & 0.75 & 0.86 & 0.95 & 0.78 & 0.86 & 0.92 & 0.70 & 0.80 & 0.95 & 0.75 & 0.84\\
    Semantic equivalents & 1.0 & 0.70 & 0.82 & 1.0 & 0.73 & 0.84 & 0.94 & 0.73 & 0.82 & 1.0 & 0.70 & 0.82\\
    \hline
    \textbf{Average} & \textbf{0.95} & \textbf{0.77} & \textbf{0.85} & \textbf{0.95} & \textbf{0.79} & \textbf{0.86} & \textbf{0.94} & \textbf{0.78} & \textbf{0.85} & \textbf{0.96} & \textbf{0.80} & \textbf{0.87}\\
    \hline
    \hline
    \textbf{Inconsistent Units} & \textbf{0.95} & \textbf{0.67} & \textbf{0.79} & \textbf{0.95} & \textbf{0.59} & \textbf{0.73} & \textbf{0.95} & \textbf{0.67} & \textbf{0.79} & \textbf{0.98} & \textbf{0.70} & \textbf{0.82}\\
    \hline
    \end{tabular}
    \label{table:fuzzy_matching}
    \vspace{-0.5cm}
\end{table*}
Another notable challenge is \textbf{informal pseudocoded mappings}, where a single test case may correspond to multiple values. We selected 100 representative test cases for this experiment. Each test case consists of two sets with multiple (key, value) pairs, and the goal is to map elements between these sets as accurately and comprehensively as possible. To increase complexity, we intentionally selected test cases where the sets contained different numbers of elements, creating scenarios where matching was non-trivial and recall could fluctuate.
To explore this, we conducted experiments under both strict (precision-focused) and relaxed (recall-focused) matching conditions. Examples reflecting different levels of strictness were included in the prompts, and the strictness level (e.g., strict, moderate, relaxed) was explicitly stated in the prompts. The results are presented in Figure \ref{fig:ambiguous}.

The experimental results indicate that under very strict conditions, precision can reach up to 100\%; however, recall drops significantly, even below 20\%. As the conditions are relaxed, precision slightly decreases, but recall increases substantially, reaching up to 55\%. Under the most relaxed conditions, recall rates for all the models approach 90\%. This demonstrates that our method can achieve high recall rates while maintaining a high level of precision.

\subsection{Time efficiency (RQ3)} 
In practical industrial scenarios, time consumption is an important criterion for measuring tool efficiency. Therefore, we measured the total time and the time taken at each stage of the SPAPI-Tester in the testing process. 
Given that the LLaMA model relies on local computational resources and that the processing speeds of GPT-3.5 and GPT-4o do not significantly differ in this pipeline, we report only the results for GPT-3.5.
We calculated the average time spent on testing all APIs. Additionally, we separately computed the time for the two major types of requests, i.e., PUT and GET. The results are shown in Table~\ref{table:time_cost}.

\begin{table}[t]
    \centering
    \caption{Time to generate test cases, per step (seconds). \textbf{DU} is document understanding; \textbf{RI} is retrieval information; \textbf{TSG} is test case generation; \textbf{Run} means running the test cases.}
    \begin{tabular}{|c|c|c|c|c|}
    \hline
    \textbf{Requests} & \textbf{Total} & \textbf{DU \& RI} & \textbf{TSG} & \textbf{Run}  \\
    \hline
    GET & 55.0 & 6.6 & 3.6 & 44.8 \\
    \hline
    PUT & 56.3 & 6.8 & 4.3 & 45.2 \\
    \hline
    \textbf{Average} & 55.7 & 6.7 & 4.0 & 45 \\
    \hline
    \end{tabular}
    \label{table:time_cost}
    \vspace{-0.4cm}
\end{table}

The results indicate that most of the time is consumed during the execution of test cases, with a significant portion dedicated to environment setup. SPAPI's complexity requires the appropriate configuration of embedded system environments, such as setting up the CAN bus for signal transmission. Additionally, the VV system needs to read CAN signals and complete the reading or setting of the virtual vehicle's state, which consumes a large amount of time.

The time required for PUT and GET requests is almost identical, as our approach batch-generates matching results or test cases for these requests, effectively minimizing time differences. In the full pipeline, the DSPy module, which leverages LLM-based capabilities, is called six times: once for documentation comprehension, four times for information matching, and once for test case generation. Additionally, DSPy’s retry mechanism re-calls the module if the output does not adhere to the predefined format. On average, the entire process from initial input to test case generation takes about 11 seconds, which is remarkably fast for automated API testing.

Manual vehicle API testing in the automotive industry is traditionally a time-consuming process, as it requires consideration of numerous conditions. To better understand the time demands of manual testing, we surveyed experts in our industry, including three senior engineers and one technical lead. They estimated that creating test cases for each API takes approximately 0.1 to 3 FTE workdays, with most APIs requiring about two hours. They generate 5 to 30 test cases for each API.

In contrast, our SPAPI-Tester achieves remarkable efficiency improvements. The system generates a complete set of test cases for a single API in just 11 seconds, representing a dramatic reduction in time and effort.
This substantial speedup not only reduces the time and effort required for API testing but also alleviates the traditionally high time burden associated with manual test case creation, greatly enhancing the API testing process.

\subsection{Performance on real-world industry APIs (RQ4)}
To demonstrate the capability of SPAPI-Tester in an real-world setting, we collected 193 newly developed and unverified truck APIs and their corresponding documentation from a leading truck manufacturing facility. We then employed SPAPI-Tester to conduct end-to-end automated testing, aiming to identify issues within these APIs.

SPAPI-Tester identified 23 test failures. The test report indicates that 22 test cases failed due to issues within the API implementation, and one test case failed due to an error while parsing the API documentation. On consultation with the API developers, these were determined to be legitimate bugs in the API implementation. The team has already started addressing these issues upon receiving the checking results.

In addition, this demonstrates that SPAPI-Tester not only has a high accuracy in detecting API errors but also provides detailed reports that help quickly identify the root causes of failures. Even when SPAPI-Tester was unable to generate correct code, the detailed reports can help to identify the failure causes quickly, thereby minimizing misdiagnoses. This capability significantly enhances the practical utility of SPAPI-Tester by providing precise and actionable insights.
In summary, these results underscore the robust practical applicability of SPAPI-Tester in real industrial environments.

\subsection{Performance comparison with manual testing}
To illustrate the advantages of SPAPI-Tester over manual API testing, we conducted a comparative evaluation. As described in Section IV.B, an expert team created ground truth test cases for 12 APIs. To measure the pass rate of manual testing, two additional engineers independently created test cases for these APIs. Results showed that one engineer’s test cases passed 10 APIs, while the other’s passed 11. 
Both engineers missed one or two APIs due to confusion over similar data entries. For instance, attributes like \texttt{reducedWeeklyRestsForCurrentWeek} and \texttt{regularWeeklyRestsForCurrentWeek} proved challenging for human testers to differentiate, whereas SPAPI-Tester’s LLMs handled them effortlessly. This led to an average pass rate of 87.5\% for manual testing at the API level, while SPAPI-Tester, with test cases generated by four different LLMs, achieved pass rates between 93\% and 98\%.

In terms of coverage, the average rate for manually created test cases was 82\%, with human testers occasionally skipping properties due to incomplete API documentation (e.g., missing units). SPAPI-Tester reached 85\% coverage with GPT-4o, while other models ranged between 73\% and 81\%.

To evaluate failure detection, we selected 10 APIs (5 of which contained known bugs) from the 109 APIs mentioned in Section IV.B. Both engineers identified all buggy APIs, although one created a test case that falsely flagged a correct API as erroneous, resulting in a recall rate of 100\% and a precision rate of 91\% for manual testing. Similarly, SPAPI-Tester achieved a recall rate of 100\% with a slightly lower precision of 90\%.

In summary, SPAPI-Tester consistently generates high-quality test cases, demonstrating comparable performance to manual testing in terms of pass rate, coverage, and failure detection.

\section{Discussion}

\noindent \textbf{On complete test process automation} -- Perhaps the most significant finding from this case study is that our recipe is capable of completely automating a real world test process. Put simply, SPAPI testing -- a process that currently takes 2-3 FTEs -- has effectively been substituted by SPAPI-Tester, a fully automatic pipeline. 
This success stems from combining LLMs with conventional automation, allowing SPAPI testing to proceed without human intervention. Key to this achievement is the nature of the SPAPI test process: it is well-structured, decomposable, and requires human judgment but not creativity. 
In such cases, LLMs serve as the critical link to full automation by systematically replacing manual steps. Maintaining the existing process structure further aids automation in two ways. First, it defines clear, verifiable steps where LLMs can be applied.
Second, preserving the status quo ensures that automation is achievable without imposing possibly unreasonable costs of changing the test process -- an observation that is crucial for real world application.

\noindent \textbf{On the generality of LLMs as problem solvers} -- Preserving the design of the process no doubt identifies discrete tasks where LLMs can be used. However, the clear enabler for complete automation is that the LLM automates all manual tasks with little practical regard to the actual nature of the task. 
Alternative automation methods exist, such as using fuzzy matching for inconsistent key-value mappings or a formal language to specify cardinality in key-value relationships. However, LLMs, as general problem solvers, eliminate the need for multiple specialized solutions, simplifying real-world implementations. 
While there is a cost to recast an LLM to solve a specific problem -- like defining prompts or signatures -- the cost turns out to be manageable.

\noindent \textbf{On implications on dependent processes} -- If SPAPI testing can be fully automated, its impact on adjacent processes becomes a natural consideration. API implementation directly precedes SPAPI testing, while integration within user-facing subsystems follows it. Given SPAPI’s simplicity, LLMs could potentially automate these dependent processes, extending automation across much of the development lifecycle—an important step for in-vehicle software engineering. Further, automating SPAPI-dependent applications could create a cascade of fully automated lifecycles, reshaping automotive software development.
While promising, this vision comes with challenges. Our results demonstrate LLMs’ ability to automate well-defined tasks and connect dependent processes, but also highlight the effort required to adapt them for specific, verifiable problems. These insights encourage further exploration toward realizing this ambitious potential.

\noindent \textbf{On the transferability of this recipe} -- We may have showcased completely automatic testing of an in-vehicle embedded software application, but it is clear that many of our observations and findings are transferable.
Our proposed criteria for automation—a decomposable process with steps requiring judgment but not creativity—can extend to other domains. Additionally, our approach involves six distinct LLM interactions: three align with general API testing workflows, while the others, though tailored to automotive scenarios, require minimal adaptation for different contexts. For example, applying this method to another vehicle manufacturer would take roughly one full workday (1 FTE). Certain aspects may also benefit web server testing. 
Finally, our recipe of largely preserving a test process and using LLMs to verifiably automate discrete manual steps is transferable to any test process that meets the criteria we propose. 

\section{Related Work}

Existing research on API testing mainly focus on black-box and white-box testing, depending on whether the source code of the API is accessible~\cite{golmohammadi2023testing}. White-box testing typically involves generating test cases to thoroughly test the logic within the code~\cite{zhang2019resource}\cite{ zhang2021resource}. For example, EvoMaster~\cite{arcuri2020automated} uses the Many Independent Objective (MIO) evolutionary algorithm to optimize multiple metrics simultaneously, such as line coverage, branch coverage, HTTP status coverage, and the number of errors. Building on this, some studies have employed additional tools for code instrumentation, such as JVM~\cite{arcuri2018test}\cite{arcuri2019restful} and NodeJS programs~\cite{zhang2022javascript}\cite{moller2019model}. Atlidakis et al.~\cite{atlidakis2020pythia} calculate code coverage by pre-configuring basic block locations and use this feedback to guide test generation. 

Currently, most studies focus on black-box API testing, aiming to enhance test case coverage for more comprehensive API testing~\cite{viglianisi2020resttestgen}. Template-based methods, such as fixed test specifications and JSON schemas, are commonly used for generating accurate test cases \cite{benac2014jsongen}\cite{chakrabarti2009test}\cite{fertig2015model}\cite{arcuri2018testmio}\cite{godefroid2020intelligent}. However, these approaches struggle to capture parameter dependencies. To address this, Stallenberg et al. \cite{stallenberg2021improving} proposed a hierarchical clustering method, while Lin et al. \cite{lin2022forest} introduced a tree-based representation of parameter relationships. Martin et al. \cite{martin2020restest} further improved test diversity by integrating external knowledge bases to generate reasonable values. Despite these advancements, traditional methods often fail to achieve robust and comprehensive testing.

Recently, LLMs have emerged as a promising direction for API testing \cite{li2024application}\cite{olasehindeoptimizing}. Kim et al. \cite{kim2024leveraging} demonstrated the utility of LLMs in interpreting natural language API documentation to generate test values. Building on this, Le et al. \cite{le2024kat} proposed constructing dependency graphs from documentation to enhance test coverage. Other studies fine-tuned LLMs using Postman test cases \cite{deepika2024automating} or applied masking techniques to predict test values \cite{decrop2024you}. However, these methods face challenges in ensuring the validity and robustness of generated test cases \cite{pereira2024apitestgenie}.

However, existing methods focus solely on test case generation, which is only one part of the API testing process, and do not address the automation of the entire process. In practical applications, these methods require significant manual verification. For instance, some approaches need to retrieve relevant yet often ambiguous information from external databases. Moreover, these methods lack robustness; if the API specification is missing parameters or contains minor errors, the process may fail. Unlike previous approaches, we are the first to explore the automation of the entire API testing process, focusing on current bottlenecks in API automation and considering how to leverage LLMs to address these challenges robustly.
\section{Conclusion}
Automated API testing is a critical process in software engineering, essential for ensuring the reliability and functionality of software systems. Despite its importance, API testing is often time-consuming, labor-intensive, and prone to errors. In practical applications, API testing involves retrieving and organizing relevant documents, and writing test cases based on the organized information. Due to the fuzzy matching of information across documents, manual intervention is required, hindering the automation of the entire testing process.

In this paper, we introduced SPAPI-Tester, the first system designed for the automated testing of automotive APIs. We decomposed the API testing process into a series of steps, identifying the obstacles to automation at each stage. By leveraging LLMs, we addressed these challenges, enabling full automation of the testing workflow. The results from real-world industrial API testing demonstrate that SPAPI-Tester achieves high detection accuracy. Our comprehensive experiments show that our system is highly robust and effective.

Our system offers valuable insights for other automated API testing tasks and can be extended to web server API testing. The findings underscore the potential of LLMs to transform API testing by reducing manual effort and improving efficiency, paving the way for broader adoption and implementation in various testing environments.

\section*{Acknowledgment}
This work was partially funded by the Wallenberg AI, Autonomous Systems and Software Program (WASP), supported by the Knut and Alice Wallenberg Foundation, and the Chalmers Artificial Intelligence Research Centre (CHAIR). The authors also thank Earl T. Barr for his insightful discussions.

\bibliographystyle{ieeetr}
\bibliography{bibliography.bib}
\vspace{12pt}

\end{document}